\documentclass[12pt,preprint]{aastex}
\newcommand{\be}{\begin{equation}}
\newcommand{\ee}{\end{equation}}
\newcommand{\bea}{\begin{eqnarray}}
\newcommand{\eea}{\end{eqnarray}}

\shorttitle{Transit Timing \& Extrasolar Trojans}
\shortauthors{Ford \& Holman}
\slugcomment{submitted to ApJL}

\begin{document}

\title{Using Transit Timing Observations to Search for Trojans of Transiting Extrasolar Planets}
\author{Eric B.\ Ford\altaffilmark{1} and Matthew J. Holman}
\affil{Harvard-Smithsonian Center for Astrophysics, Mail Stop 51, 60 Garden Street, Cambridge, MA 02138}
\email{eford,mholman@cfa.harvard.edu}

\altaffiltext{1}{Hubble Fellow}

\begin{abstract}
Theoretical studies predict that Trojans are likely a frequent
byproduct of planet formation and evolution.  We examine the
sensitivity of transit timing observations for detecting Trojan
companions to transiting extrasolar planets.  We demonstrate that this
method offers the potential to detect terrestrial-mass Trojans using
existing ground-based observatories.  We compare the transit timing
variation (TTV) method with other techniques for detecting extrasolar
Trojans and outline the future prospects for this method.
\end{abstract}
\section{Introduction} 


For centuries, theories of planet formation had been
designed to explain our own Solar System, but the first
extrasolar planetary systems discovered were very different than our
own (e.g., Mayor \& Queloz 1995).  These discoveries led to the
realization that planet formation theory must be generalized to
explain a much greater diversity of planetary systems. 
Stable Trojan companions to extrasolar planets may be common.  In our
solar system, Mars, Jupiter, and Neptune each share their orbit with
asteroids orbiting near the stable (L4/L5) Lagrange points that
lead/trail the planet by $\simeq~60^{\circ}$.  
While the mass
ratios of the Trojan systems in our solar system are rather extreme
($\le7\times10^{-9}$), it is possible that extrasolar planets may have
much more massive Trojans.  Indeed, theorists have already outlined
several possible mechanisms to form Trojans with mass ratios
potentially including unity (Laughlin \& Chambers 2002; Chiang \& Lithwick 2005; Morbidelli et al.\ 2005; Thommes 2005; Cresswell \& Nelson 2006; Ford \& Gaudi 2006).  

Trojans of both Jupiter and Neptune have provided clues about our own
solar system's history (Michtchenko, Beauge \& Roig 2001; Kortenkamp,
Malhotra \& Michtchenko 2003; Chiang \& Lithwick 2005; Morbidelli et
al.\ 2005).  Similarly, the detection of extrasolar Trojans would be
useful for constraining theories of planet formation and migration.
While all the above mechanisms predict that Trojans would survive the
migration process, there are alternative models of planet migration
that predict Trojans would not survive (Rasio \& Ford 1996; Wu \& Murray 2003; 
Gaudi 2003; Ford \& Rasio 2006; Ford \& Gaudi 2006).
The detection of a Trojan companion to a short-period planet would
present a serious challenge to these mechanisms for forming ``hot
Jupiters'' and would imply that the planet in such a system was formed
via migration through a dissipative disk rather than tidal
circularization after approaching the star on a highly eccentric
orbit.  Thus, searching for extrasolar Trojans can test models of
planet formation.
%


Previously, three methods have been proposed to identify extrasolar
Trojans.  If a Trojan is sufficiently massive and has a sufficiently
large libration amplitude, then its presence could be inferred using
the deviations from a Keplerian perturbation to the stellar radial
velocity or astrometric signal caused by a single planet.  Laughlin \&
Chambers (2002) have shown that two comparable mass giant planets
occupying a 1:1 mean motion resonance would typically have strong
planet-planet gravitational interactions on a secular timescale.
However, these signatures may not be unique: a reanalysis of the RV
observations of HD 128311 and HD 82943 have shown that both of the
current data sets are also consistent with a pair of planets in a 1:1
mean motion resonance (Gozdziewski \& Konacki 2006), as well as the
originally published orbital solutions.

If a Trojan transits its parent star, then
photometric or spectroscopic monitoring of stars with transiting
planets (particularly at times offset from the planet transit by
$\sim~P$/6) may reveal the Trojan transit via the decrease in stellar
flux (Rowe et al.\ 2000) or anomalous RV excursions due to the
Rossiter-McLaughlin effect (Gaudi \& Winn 2006).  Both these methods
are more sensitive to large Trojans.  While ground-based observations are
not sensitive to Earth-sized planets, space observations could detect
such Trojans.  However, it is not guaranteed that a Trojan will transit its
parent star, as it may have a significant inclination (e.g.,
Morbidelli et al.~ 2005).  Further, since the libration period can be quite
large, long-term monitoring would be required to ensure detection.

Ford \& Gaudi (2006) proposed a method for detecting a Trojan
companion based on combining radial velocity observations and
photometric observations of a transiting planet.  Even if the Trojan
itself were not transiting, it could reveal it's presence via a time
lag between the radial velocity null and the time of central transit.
Existing observations already place significant (99.9\%) upper limits
on the mass of Trojan companions to HD 209458b and HD 149026b of
$\simeq~13M_\oplus$ and $\simeq~25M_\oplus$.

Here, we present another method for detecting Trojan companions to
extrasolar planets using only photometric observations of transiting
extrasolar planets.
Once a transiting planet has been identified, higher precision
follow-up observations and modeling can precisely measure the mid-time
of subsequent transits (currently with a precision $\sim$10s; e.g.,
Agol \& Steffen 2006; Holman et al.\ 2006; Winn et al.\ 2006).  If the
star and the transiting giant planet are the only bodies in the
system, then the transits will be strictly periodic, i.e., $t_i = t_0
+ i\times P_s + \delta t_i$, where $t_i$ is the time of the $i$th
transit, $P_s$ is the transiting planet's sidereal orbital period, and
any transit timing variations ($\delta t_i$) are due to measurement
error.  However, if an additional planet orbits the star, then the
times of the giant planet's transits will be affected (Miralda-Escude
2002; Holman \& Murray 2005; Agol et al.\ 2006; Heyl \& Gladman 2006).
%
%
By analyzing the deviations of the observed TTVs from
a strictly periodic model ($\delta t_i$), astronomers can search for
additional planets orbiting the star. Here, we show that a sub-Earth
mass Trojan planet could induce a transit timing signal that is easily
measurable using existing ground-based observatories.

\section{Observational Constraints on Trojans}
We consider a three body system and denote the stellar mass
($m_{\star}$), the planet mass ($m_p$), and the Trojan mass ($m_T$).
We refer to all bodies librating about the L4 or L5 fixed point of a
planet as ``Trojans''.
If there are no
other massive bodies in the system, then the L4/L5 fixed points are
stable for circular orbits if the ratio,
$\mu=(m_p+m_T)/(m_\star+m_p+m_T)$, is less than a critical threshold
$\mu_c$, where $0.03812\le\mu_{c}\le0.03852$ and $\mu_c$ depends on
the ratio, $\epsilon\equiv m_T/(m_p+m_T)$ (Laughlin \& Chambers 2002).
If the planet and Trojan have equal eccentricities and the Trojan
resides exactly at the L4/L5 fixed point, then the transit timing
signature for the primary planet would be indistinguishable from a
similar system without a Trojan.  
More generally, for a Trojan on an orbit that is librating about the
L4/L5 fixed point, the times of the primary planet's transits will
deviate from being strictly periodic.  Here, we focus our attention on
Trojans that undergo small librations about the L4/L5 fixed points and
are significantly less massive than the currently known planet.

The libration can be approximated as a linear superposition of two
epicyclic motions.  The star-Trojan separation can oscillate
about the semi-major axis of the planet ($a_p$) with the amplitude,
$\delta a \ll \mu^{1/2} a_p$ on a timescale $\tau_{\rm fast}\simeq
P_s$, and the guiding center of the Trojan can oscillate on a longer
timescale, $\tau_{\rm slow}\simeq P_s\sqrt{27/(4\mu)}$ (Murray \&
Dermott 2000).  The timescale of the libration of the guiding center
makes this motion most readily detectible by transit timing
observations.

For a transiting planet, both $P_s$ and each $t_i$ can be measured
precisely using photometry alone.  Considering a series of continuous
photometric observations with uncorrelated Gaussian uncertainties of
magnitude $\sigma_{ph}$, taken at a rate $\Gamma$ around a single
transit, the mid-transit time can be measured with an accuracy of
$\sigma_{t_i}\simeq~\sqrt{t_e/2\Gamma}\sigma_{ph}\rho^{-2}$, where
$t_e$ is the duration of ingress/egress and $\rho$ is the ratio of the
planet radius to stellar radius (Ford \& Gaudi 2006).  For typical
parameters (e.g., $\sigma_{ph}\sim~10^{-3}$), $t_i$ can be measured to
$\simeq10$s (e.g., Brown et al.\ 2001; Holman et al.\ 2006).  The
period can be measured much more accurately than $t_i$, from
observations of multiple transits separated by many orbits.
For small amplitude libration about L4/L5 and circular orbits, the
transit timing perturbation is given by $\delta t_i \simeq \epsilon
P_s \Delta M(t_i) / (2\pi)$, where $\Delta M(t_i)$ is the angular
displacement of the Trojan from L4/L5 at the time of the $i$th
transit.  The TTVs can be modeled by a sinusoid,
$\delta t_i = K_{\rm tt} \sin\left(2\pi\left(t-t_0\right)/P_{\rm TTV}+\phi\right)$,
where 
$K_{\rm tt}$ is the amplitude of the transit timing variations
and $P_{\rm TTV}\sim\tau_{\rm slow}$.
If the dominant periodicity of the transit timing variations ($P_{\rm
TTV}$) is well determined, then the remaining parameters can be
determined via linear least squares fitting to the observed transit
times.  
The transit timing variations will have an amplitude
\be
K_{\rm tt} \simeq 60s \left(\frac{P}{4d}\right) \left(\frac{m_T}{m_\oplus}\right) \left(\frac{0.5 M_{\rm Jup}}{m_p+m_T}\right) \left(\frac{K_{\Delta M}}{10^\circ}\right),  \\
\ee
where $K_{\Delta M}$ is the amplitude of the Trojan's angular displacement from the Lagrange point.  For small amplitude libration, $K_{\Delta M}\simeq\mathrm{max}\left|\Delta M\right|$ and $\mathrm{rms}(\delta t_i)\simeq K_{\rm tt} / \sqrt{2}$ (see Fig. 1)
Libration amplitudes of $K_{\Delta M}\sim~5-25^\circ$ are common for
Trojans orbiting near the Sun-Jupiter Lagrange points (Murray \&
Dermott 2000).  

The Lomb-Scargle periodogram can be easily adapted to efficiently scan
a range of putative periods and identify any significant periodicities
(Cumming 2004).  If we assume that there are many ($N_{\rm tt}$)
transit timing observations with uncorrelated Gaussian uncertainties
$\sigma_{t_i}=\sigma_{tt}$, that the transit timing observations are
evenly distributed, and the duration of observations ($T_{\rm obs}$)
is greater than than $P_{\rm TTV}$, then a periodogram-style analysis
results in a 50\% chance of detecting a Trojan if
$K_{tt}\ge K_{1/2} \simeq \sigma_{tt}\left(\frac{4}{N_{\rm tt}}
\log\left[T_{\rm obs} / \left(2 F P_s\right) \right] \right)^{1/2}$
(Cumming 2004), where $F$ is the false alarm probability, which we set
to $10^{-3}$.  For $N_{\rm tt} = T_{\rm obs}/P_s = 40$, $K_{1/2}
\simeq \sigma_{tt}$, so sub-Earth-mass Trojans could be readily
detected.  We note that all published transit timing data sets have
$N_{\rm tt}<20$, which results in a significantly reduced sensitvity,
if $P_{\rm TTV}$ is unknown {\em a priori}.  In this small-$N_{\rm
tt}$ regime, a simple $\chi^2$ test of the null hypothesis ($\delta
t_i=0$) is more sensitive for detecting transit timing variations.
However, if only a single periodicity (e.g., $\tau_{\rm slow}$) is to
be tested, then even a modest number of observations can be quite
sensitive (e.g., $K_{1/2}\simeq 2.5\sigma_{\rm tt}$ even for $N_{\rm
tt}=13$).

Once a Trojan has been detected, a Fisher information analysis (e.g.,
Gaudi \& Winn 2006) reveals that the uncertainty in $K_{\rm tt}$ will
approach 
$\sigma_{K_{\rm tt}} = \sqrt{4/N_{tt}} \sigma_{tt}$.
If a Trojan were present, then the uncertainty in
$K_{\rm tt}$ would set the uncertainty in the measurement of the
mass of the Trojan to be
\be
\label{EqnMass}
\frac{\sigma_{m_T}}{m_\oplus}  \simeq  \frac{0.5}{\sqrt{N_{\rm tt}}} \left(\frac{\sigma_{\rm tt}}{10s}\right) \left(\frac{4d}{P}\right) \left(\frac{m_p}{0.5 M_{\rm Jup}}\right) \left(\frac{10^\circ}{K_{\Delta M}}\right)  \\
\ee
Thus, transit timing observations can be very sensitive to
sub-Earth-mass Trojan companions.  However, due to the degeneracy
between $m_T$ and $K_{\Delta M}$, the amplitude of transit timing
detections due to a Trojan would not provide a strong upper limit on
the Trojan mass.  For Trojans with significant libration amplitudes,
this degeneracy could be resolved by combining the amplitude with the
measured $P_{\mathrm{TTV}}$ (see Fig.\ 1 center).
For Trojans with small libration amplitudes, the measured $P_{\mathrm{TTV}}$
will provide an upper limit for the libration amplitude and
hence a lower limit to the Trojan-planet mass ratio.  Regardless, the
transit timing observations could be used to predict the time of
transit of a Trojan and targeted photometric or spectroscopic
follow-up observations could place an upper limit on the radius of the
Trojan (e.g., Rowe et al.\ 2006; Gaudi \& Winn 2006).  Combining such
observations with planetary structure models (e.g., Valencia et al.\
2007) could provide an upper limit on the Trojan mass and hence a
lower limit for the libration amplitude.

\section{Discussion}


One long-term goal of immense scientific and public interest is to
discover and study rocky planets, and eventually terrestrial planets
that have masses, sizes, orbits, atmospheres, and perhaps even surface
conditions similar to those of the Earth.
Previous studies have demonstrated that the magnitude and timescale of
transit timing variations due to Earth-mass planets are readily
detectable if they orbit near an interior or exterior low-order mean
motion resonance (Holman \& Murray 2005; Agol et al.\ 2005).  The TTV
method is quite sensitive to planets near resonances (e.g., the
planetary systems GJ 876, HD 128311, HD 73526, 55 Cnc, and HD 82943;
Butler et al.\ 2006) that are particularly valuable for studying
orbital dynamics and planet formation (Lee \& Peale 2002) and
challenging for radial velocity and astrometric searches.  The TTV
method would also be able to confirm some planet candidates (likely to
be identified by future transit searches) by detecting the orbital
interactions of the planets, similar to the methods used for
confirming the planets around PSR1257+12 (Rasio 1992; Malhotra 1993)
and PSR1620-26 (Ford et al.\ 2000).  This could prove particularly
valuable for planet candidates that have small masses and/or orbit
faint stars, so that radial velocity confirmation is impractical
(e.g., most of the 16 transiting planet candidates orbiting faint
stars recently published by Sahu et al.\ 2006, and the many transiting
planets expected to be found by future space missions). \hfill

It would be extremely exciting to detect a {\em transiting} Earth-mass
planet.  Such a detection would enable follow-up observations to study
the physical properties of the planet, such as the planet's radius and
density (Brown et al.\ 2001, Sato et al.\ 2005, Charbonneau et al.\
2006), the atmospheric composition (Charbonneau et al.\ 2002, Deming
et al.\ 2005, Bozorgnia et al.\ 2006), and possibly even ``resolve''
surface/atmospheric features (Ford et al.\ 2001; Harrington et al.\
2006; Gaidos et al.\ 2006).

We have demonstrated that a sub-Earth-mass Trojan planet could also
result in a transit timing signal that can be readily measured with
ground based observatories.  Since the orbital planes are likely
nearly aligned, the fact that a giant planet is already known to
transit the star increases the odds that other planets orbiting that
star will also transit (Holman \& Murray 2005).  Thus, the transit
timing method is particularly good at searching for transiting
Earth-like Trojan planets that would enable extremely interesting
follow-up observations. Our technique could be applied to search for
terrestrial-mass Trojans of giant planets orbiting in the habitable
zone of their stars (Schwarz et al.\ 2005), particularly for low mass
stars where the habitable zone can be $\simeq~0.015$~AU away from the
star.  Once transitting terrestrial mass planets are discovered, this
technique could be extended to search for extrasolar Trojans with
asteroid-like masses.

While \S2 and previous work have emphasized the sensitivity of transit
timing observations, we caution that solving the inverse problem of
determing planet properties from transit timing observations is likely
to pose a significant challenge and be more difficult than
interpreting other types of extrasolar planet observations.  For
example, in the radial velocity method, the dominant periodicity in
the observed time series is readily identified with the orbital period
of a massive companion and the amplitude of the variations is
proportional to the mass of the companion (Konacki \& Maciejewski
1999).  However, in TTV data, the dominant periodicity could be due to
any one of several physical effects (see Fig.\ 2), including the
reflex motion of the star due to the second planet (with a period
equal to the orbital period of the second planet), the long-term
mutual gravitational perturbations between the planets (with a period
much longer than either orbital period), the short-term gravitational
perturbations on the orbit of the transiting giant planet (on an
intermediate timescale), or the light travel time due to a distant
companion (e.g., Borkovitz et al.\ 2003; Heyl \& Gladman 2006).
Therefore, even once a periodicity has been identified, it is not
obvious what physical effect is causing the periodicity.
Further, TTV signatures are more complex than the signatures of other
dynamical detection techniques.  For example, radial velocity
observations of a multiple planet system can often be modeled by the
linear superposition of multiple Keplerian orbits (Butler et al.\
2006; Ford et al.\ 2006).  However, for transit timing observations,
the signal is often dominated by the deviations from such a simplified
model.
Therefore, it is necessary to perform n-body simulations to accurately
calculate the TTV signature of each possible model (Holman \& Murray
2005; Agol et al.\ 2005; Steffen \& Agol 2005).  Given the
computational requirements of each n-body integration, practical
algorithms must explore the high-dimensional ($\simeq7\times~N_{pl}$)
parameter space very efficiently and rapidly converge on all physical
models consistent with the observations.

A Trojan companion with small libration amplitude will induce a TTV
signature that can be well approximated by a single sinusoid (Fig.\ 2,
top).  This contrasts with the TTV perturbations due to a planet near
a different resonance (Fig.\ 2, middle and bottom).  A moon could also
result in a nearly sinusoidal TTV signature, but on a timescale this
is typically much shorter than $\tau_{\mathrm{slow}}$ (due to
dynamical stability constraints).  TTV perturbations with a timescale
near $\tau_{\mathrm{slow}}$ and due to non-resonant planets will have
small amplitude, unless the outer planet is quite massive and
potentially detectable by other methods.  Therefore, we suggest that a
large amplitude sinusoidal signal with a period near
$\tau_{\mathrm{slow}}$ might allow Trojans to be uniquely identified.
We suggest future investigations to test this conjecture.

We caution that the TTV signature of an extrasolar Trojan could also
be non-sinusoidal.  For example, Trojans with large libration
amplitudes can become significantly non-sinusoidal (reducing the rms
TTV by upto $\sim40\%$).  A Trojan planet in a horseshoe-shaped orbit
would produce much larger TTV perturbations with a very different
shape.  If the primary and Trojan planets have different
eccentricities, then there will be additional longer term
periodicities in the TTV signal due to secular perturbations (Fig.\ 1,
right).  If there are Trojans at both L4 and L5, then the transit
timing signature could be approximated as the sum of two such signals
(similar frequency, but different amplitudes and phases).  Similarly,
swarms of Trojan companions librating about L4 and L5 could be modeled
as the superposition of many such signals, provided that their mutual
interactions are negligible.  Additional planets could also perturb
the time of central transit (Holman \& Murray 2005; Agol et al.\ 2005)
such that the offset will vary from transit to transit.  Therefore,
many transits should be observed to verify that any observed offsets
are not due to perturbations by a more distant giant planet.

The interpretation of actual TTV observations will be further
complicated by constrained sampling (observations only possible during
transit), incomplete sampling (due to available telescope time, and
weather; Agol \& Steffen 2006) and measurement errors, all of which
increases the uncertainties in the number, masses and orbits of
planets.  These limitations underscore the need for powerful
statistical methods to interpret TTV observations.  In cases where multiple
orbital models are consistent with TTV data, additional
observational constraints (e.g., radial velocities,
secondary transit, changing transit duration due
to inclination librations) could help identify the correct model.  We
encourage further research in such methods, so that ongoing TTV
observations can be appropriately analyzed.

\acknowledgments
We thank Eric Agol, Dan Fabrycky, Scott Gaudi, Jason Steffen, and Josh Winn, for helpful comments.  
Support for EBF was provided by NASA through Hubble Fellowship grant
HST-HF-01195.01A awarded by the Space Telescope Science Institute,
which is operated by the Association of Universities for Research in
Astronomy, Inc., for NASA, under contract NAS 5-26555.  
MJH acknowledges support for this work NASA Origins grant
NG06GH69G.

\newpage

\begin{figure}[htbp]
\plotone{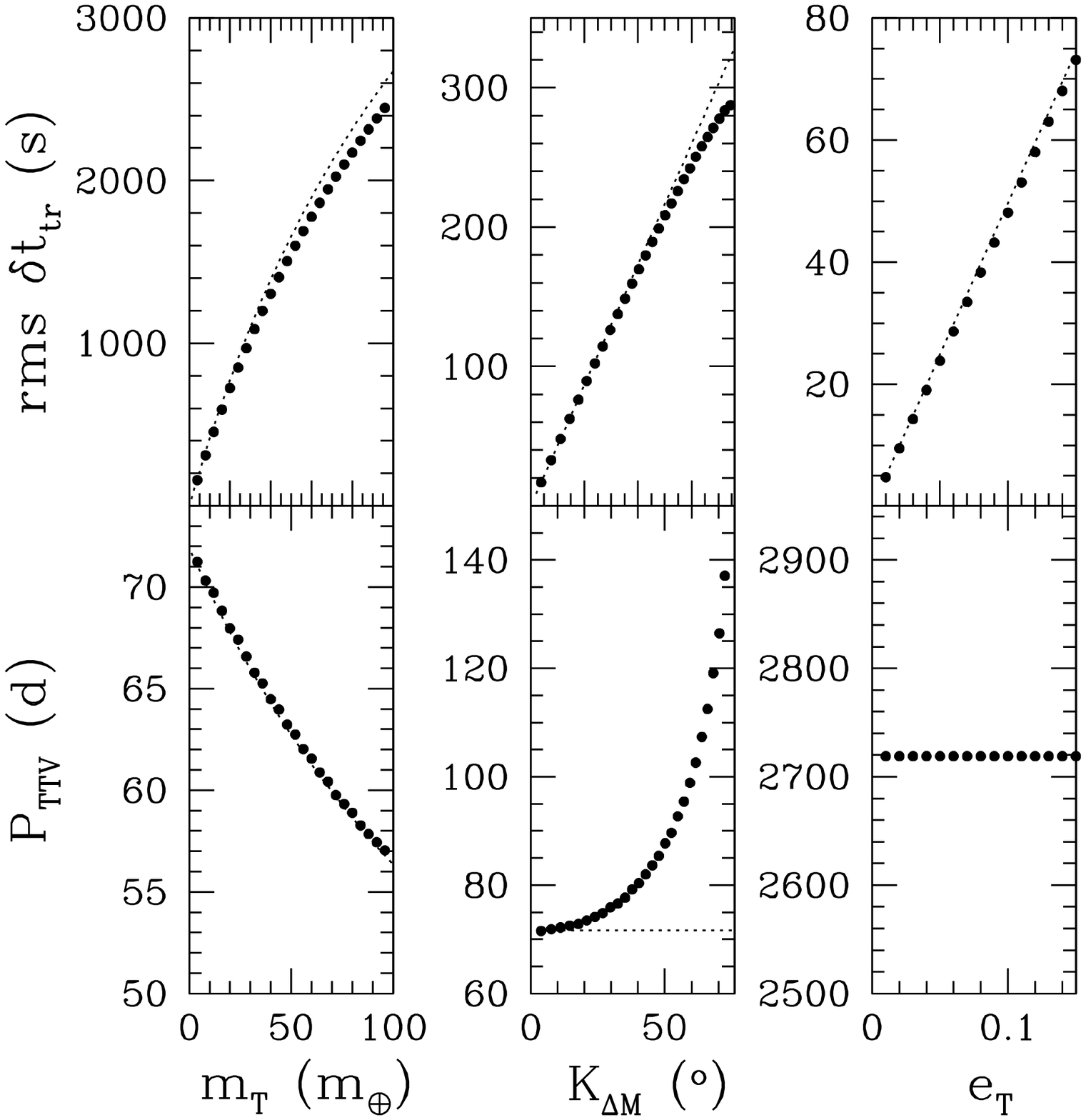} 
\caption{
Transit Timing Signatures:
We show the root mean square
deviations of the TTVs from a strict periodicity (top) and
the period of the transit timing variations (bottom)
based on direct n-body integrations of a system with 
a $0.5M_{\rm Jup}$ planet and a Trojan orbiting a $M_\odot$ star.  
Dotted lines are analytic expressions from \S2.
(Left) The planet and the Trojan companion with mass $m_T$ are
initially placed on circular orbits with a mean orbital separation of
0.05AU and $\Delta M_T=10^{\circ}$.
(Center) As before, but as a function of $K_{\Delta M}$, the amplitude of the angular displacement from the Lagrange point, for a fixed Trojan mass of $1M_{\oplus}$.  
(Right) As before, but as a function of initial eccentricity of the Trojan for a fixed Trojan mass of $1M_{\oplus}$ and initial $\Delta M_T=0^{\circ}$.  The dotted curve shows the analytic model, $\mathrm{rms}(\delta t_{tr}) = \epsilon P_s e_T /(\pi\sqrt{2})$.
\label{fig_ttv_amp_mass} }
\end{figure}

\begin{figure}[htbp]
\plotone{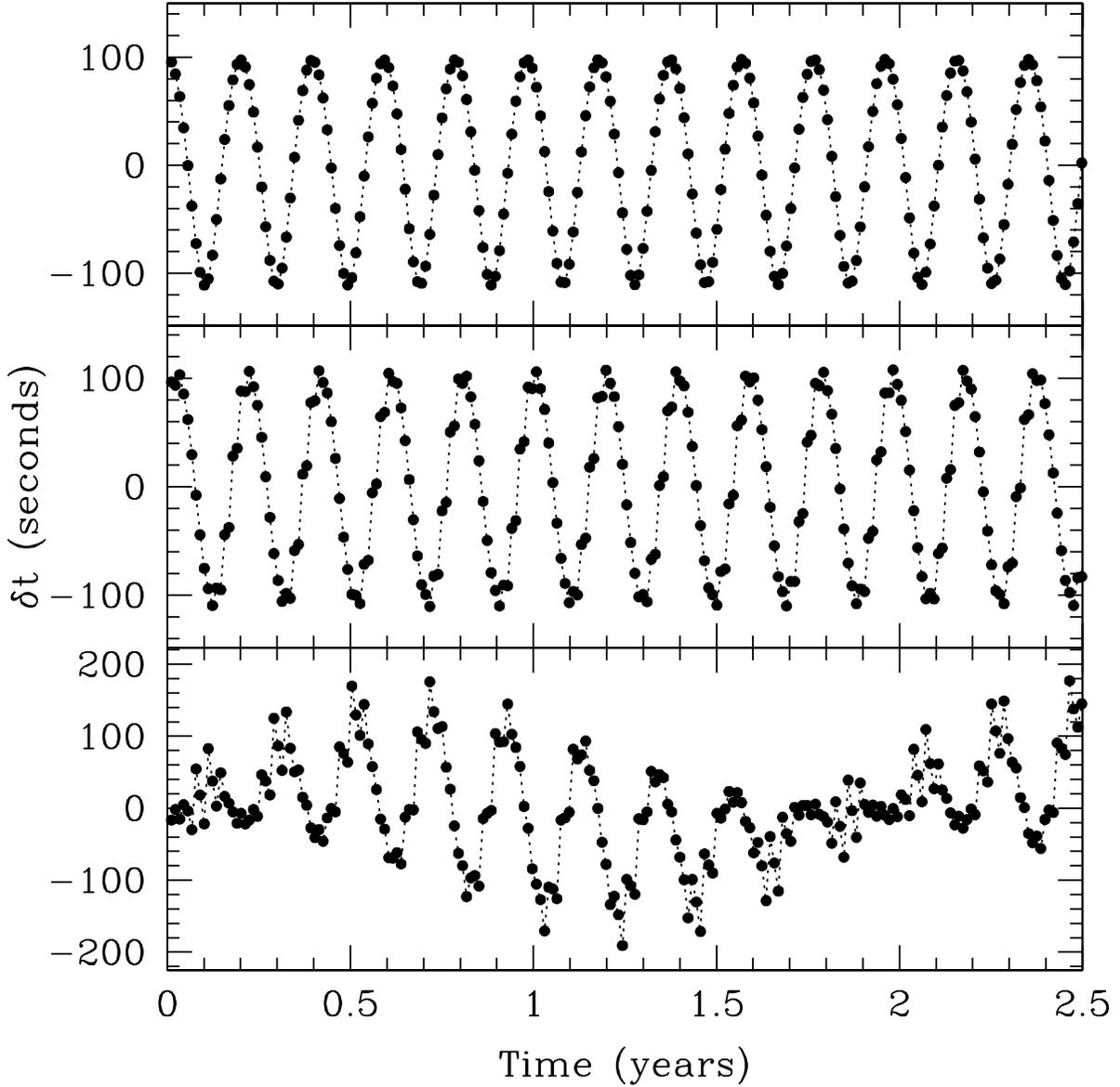} 
\caption{
%
Similar Transit Timing Signatures due to Very Different
Perturbing Planets: We plot the TTV residuals (disks) versus time for
three hypothetical planetary systems.  (The dotted lines merely guide
the eye.)  Each contains a typical transiting giant planet (0.5
Jupiter masses, orbital period of $4.09$days) and a second
planet.  The perturbations are due to: top) a $1M_{\oplus}$
Trojan companion, middle) the
perturbations are due to a $28M_{\oplus}$ (or 0.3 Saturn-mass) planet
with a period of $\simeq~8.7$days (outside the 2:1 mean-motion
resonance), and bottom) a
$\simeq~4.8$ Earth-mass planet with a period of $\simeq~5.91$days
(inside the 3:2 mean-motion resonance).  Each planetary systems
results in a TTV signature that has a dominant periodicity of
$71.40$days and a root-mean-square amplitude of $73.4$s (based 1000
transits).  Thus, interpreting TTV observations will require combining
dynamical analyses with advanced statistical methods.
%
%
\label{fig_ttv_ex_comp}  }
\end{figure}

\end{document}